\documentclass[12pt]{article}
\usepackage{amssymb,epsf}
\newcommand{\Purple}[1]{#1}
\newcommand{\Red}[1]{#1}
\newcommand{\Blue}[1]{#1}
\newcommand{\ForestGreen}[1]{#1}
\newcommand{\ft}[2]{{\textstyle\frac{#1}{#2}}}

\csname @addtoreset\endcsname{equation}{section}

\begin{document}

\begin{titlepage}
\begin{flushright}
UG/00-18\\
SU-ITP-00/31\\
KUL-TF-00/29\\
hep-th/0012110
\end{flushright}
\vspace{.5cm}
\begin{center}
\baselineskip=16pt
{\LARGE    Supersymmetry of RS bulk and brane  $^{\dagger}$
}\\
\vfill
{\large Eric Bergshoeff $^{1}$, Renata Kallosh,$^{2}$, 
and Antoine Van Proeyen $^{3}$  } \\
\vfill
{\small
$^1$ Institute for Theoretical Physics, Nijenborgh 4,
9747 AG Groningen, The Netherlands \\[3mm]
$^2$ Department of Physics, Stanford University, Stanford,
CA 94305, USA
\\[3mm]
${}^3$ Instituut voor Theoretische Fysica, Katholieke
 Universiteit Leuven,\\
Celestijnenlaan 200D B-3001 Leuven, Belgium
}
\end{center}
\vfill
\begin{center}
{\bf Abstract}
\end{center}
{\small We review the construction of actions with supersymmetry on spaces
with a domain wall. The latter objects act as sources inducing a jump in
the gauge coupling constant. Despite these singularities, supersymmetry
can be formulated, maintaining its role as a square root of translations
in this singular space. The setup is designed for the application in five
dimensions related to the Randall--Sundrum (RS) scenario. The space has
two domain walls. We discuss the solutions of the theory with fixed
scalars and full preserved supersymmetry, in which case one of the branes
can be pushed to infinity, and solutions where half of the
supersymmetries are preserved.\vspace{2mm} \vfill \hrule width 3.cm} To
be published in the proceedings of the NATO advanced research workshop
\emph{Noncommutative structures in mathematics and physics }in Kiev and
in the proceedings of the EC-RTN workshop \emph{The quantum structure of
spacetime and the geometric nature of fundamental interactions } in
Berlin. Talks given by A.V.P.
\end{titlepage}

\section{Introduction}

It is not obvious how supersymmetry can be implemented in a space with
domain walls. The wall is at a fixed place and its presence seems to lead
to a breaking of translations orthogonal to the plane. Supersymmetry,
being the square root of translations, seems rather difficult to realize
in this context. It is interesting to see how this obstacle has been
avoided in~\cite{vpr-susyd5sing}, which we summarize here.

The work is mostly motivated by the Randall--Sundrum (RS)
scenarios~\cite{vpr-RSIandII}. The simplest form of the situation that is
under investigation consists of a 3-brane in a 5-dimensional bulk. The
solution can be generalized e.g. to 8-branes in $D=10$, but the full
implementation of that situation is still under investigation.

When the RS scenarios appeared, supersymmetrisation was soon
investigated. After initial attempts, it was found that no smooth
supersymmetric RS single-brane scenario is possible~\cite{vpr-KLandBCII}. This
scenario with one brane was put forward as an alternative to
compactification.

This lead us to the original RS setup with two branes. The 2-brane
scenario has a compactified fifth dimension, $x_5\simeq x_5+2\tilde x_5$,
with two branes fixed at $x_5=0$ and $x_5=\tilde x_5$.
\begin{figure}
\begin{center}
\leavevmode \epsfxsize=6cm
 \epsfbox{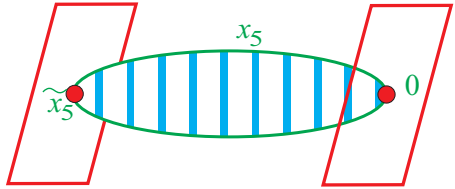}
\caption{\it Two-brane scenario. The fifth dimension is a circle with
branes at opposite ends and a $\mathbb{Z}_2$ identification of points
symmetric w.r.t. $x_5=0$. \label{vpr-fig:2brane}}
\end{center}
\end{figure}
There is moreover an orbifold condition relating points $x_5$ and $-x_5$.
Thus, the five-dimensional manifold has the form ${\bf M}= {\bf M}_4\times
{S^1\over \mathbb{Z}_2}$. This is similar to the
Ho\v{r}ava--Witten~\cite{vpr-HoravaWitten} scenario. The latter one embeds
10-dimensional manifolds in an 11-dimensional space. They obtain the
supersymmetry by a  cancellation between anomalies of the bulk theory and
a non-invariance of the classical brane action. Lukas, Ovrut, Stelle and
Waldram~\cite{vpr-LOSW} reduced this on a Calabi--Yau manifold to five
dimensions, and further developed this setup in five dimensions. Further
steps have been taken
by~\cite{vpr-Pomarol,vpr-Bagger,vpr-Polish,vpr-Zucker}.
In~\cite{vpr-Bagger,vpr-Zucker} the gauge coupling constant does not
change when crossing the branes, while in~\cite{vpr-Pomarol,vpr-Polish}
this coupling constant changes sign. In that respect, our approach is most
close to the latter. In these papers, the action in the bulk is modified,
such that it is not supersymmetric any more by itself, but the
non-invariance is compensated by the brane action to obtain invariance of
the total action. We~\cite{vpr-susyd5sing} obtain separate invariance of
bulk and brane action.

The first part of this report will treat the construction of the action
with local supersymmetry on the singular space. In that part, we will
show how the bulk and brane action are separately invariant under
supersymmetry. The supersymmetry that we are considering is the one with
8 real components, i.e.\ minimal (${\cal N}=2$) supersymmetry in 5
dimensions. The algebra is preserved despite the discontinuity. The
second part treats background solutions. The Killing spinors are
discussed. There are solutions with fixed scalars and 8 Killing spinors,
and solutions of $1/2$ supersymmetry, i.e.\ with 4 Killing spinors.
Finally a summary is given, discussing open issues.
\section{The action for bulk and brane}
The construction of the action involves three steps. First, we consider
the bulk action. That is the action of supergravity in $D=5$ with matter
couplings. A quite general action has been given
in~\cite{vpr-AnnaGianguido} based on the general methods developed in 4
dimensions in~\cite{vpr-Andrianopoli}. But it may not be excluded that
further generalizations are possible~\cite{vpr-qgend5matter}. We will
restrict ourselves to the couplings of vector multiplets, for which the
general couplings were found in~\cite{vpr-GST}. One can separate the
ungauged part, and the part dependent on a gauge coupling constant $g$.
We will consider only the gauging of a $\mathop{\rm {}U}(1)$ $R$-symmetry
group.

In the second step, the gauge coupling constant $g$ is replaced by a
field $G(x)$. A Lagrange multiplier field, a $(D-1)$-form (4-form for our
application), is introduced, whose field equation imposes the constancy of
$G(x)$ such that effectively it is still a constant.

The third step introduces the brane action. That action has extra terms
for the Lagrange multiplier $(D-1)$-form, which allows $G(x)$ to vary
crossing the brane. We will show how every step preserves the
supersymmetry!

Before embarking on that programme, we want to repeat the fundamental
algebraic relation between the cosmological constant and the gauge
coupling constant of $R$-symmetry. The super-anti-de~Sitter algebra for
${\cal N}=2$ in $D=5$ is $\mathop{\rm SU}(2,2|1)$. It involves the
anti-de~Sitter algebra $\mathop{\rm SO}(4,2)\simeq \mathop{\rm SU}(2,2)$
with translations $P_a$ and Lorentz rotations $M_{ab}$, the
supersymmetries $Q^i$, with $i=1,2$, a symplectic Majorana spinor, and a
$\mathop{\rm {}U}(1)$ generator as $R$-symmetry. The most characteristic
(anti)commutator relations are
\begin{eqnarray}
 \left\{ \Red{Q^i},\Red{Q^ j}\right\}  & = &
 \ft12\varepsilon ^{ij}\gamma _{a } \Red{P^a} +{\rm i} \Blue{g Q^{ij}}
 \gamma ^{ab}\Red{M_{ab}}
 +{\rm i}\varepsilon ^{ij}\Red{U}\,,\nonumber\\
\left[ \Red{U},\Red{Q^i} \right]&=& \Blue{g Q^i{}_j}\,
\Red{Q^j}\,,\nonumber\\
\left[ \Red{P_a},\Red{P_b}\right] &=& \Blue{g^ 2Q^i{}_jQ^j{}_i} \, \Red{M_{ab}}\,,\nonumber\\
\left[ \Red{P_a},\Red{Q^i}\right] &=& {\rm i} \gamma _a \Blue{g
Q^i{}_j}\Red{Q^j}\,.
 \label{vpr-undeformedAlgebra}
\end{eqnarray}
$\Blue{Q_{ij}}$ satisfies
\begin{eqnarray}
  &&\Blue{Q_{ij}}=\Blue{Q_{ji}}\,,\qquad\Blue{Q^i{}_j}\equiv \varepsilon
  ^{ik}\Blue{Q_{kj}}={\rm i}\left( \Blue{q_1} \sigma _1 + \Blue{q_2} \sigma _2
+\Blue{q_3 }\sigma _3\right)\,,\nonumber\\
&&q_1,\, q_2\,, q_3\in \mathbb{R}\,,\qquad (q_1)^2+(q_2)^2+(q_3)^2=1\,.
 \label{vpr-matrixQ}
\end{eqnarray}
This matrix determines the embedding of $\mathop{\rm {}U}(1)$ in the
automorphism group of the supersymmetries $\mathop{\rm SU}(2)$. This
choice is not physically relevant in itself.  The second of the
commutators in~(\ref{vpr-undeformedAlgebra}) implies that $g$ is the
coupling constant of $R$-symmetry. But the third equation says that $g^2$
determines the curvature of spacetime, i.e.\ it determines the
cosmological constant. This fact is the cornerstone of the situation that
we describe. The gauge coupling and the cosmological constant are
related. However, one can change the coupling constant from $+g$ to $-g$,
not affecting the cosmological constant. That is what will happen going
through the branes. This jump in the sign of $g$ will thus occur together
with the action of the $\mathbb{Z}_2$. This $\mathbb{Z}_2$ acts on the
fields, which therefore live on an orbifold. One can distinguish odd and
even fields. The circle condition on the fields and the orbifold
condition are then
\begin{eqnarray}
&&\Phi (\ForestGreen{x^5})=\Phi(\ForestGreen{x^5 +2\tilde x^5})\,,\nonumber\\
&&  \Phi _{\rm even}(-\ForestGreen{x^5})=\Phi _{\rm
even}(\ForestGreen{x^5})\,,\qquad \Phi _{\rm odd}(-\ForestGreen{x^5})=
  -\Phi _{\rm odd}(\ForestGreen{x^5})\,.
 \label{vpr-circleorbif}
\end{eqnarray}
These conditions imply that odd fields vanish on the branes: at
$\ForestGreen{x^5=0}$ and at $\ForestGreen{x^5=\tilde x^5}$.

Also the supersymmetries split. Half of them are even, and half are odd.
Therefore, on the brane one has 4 supersymmetries, i.e.\ ${\cal N}=1$ in 4
dimensions. This splitting of the fermions requires a projection matrix
in $\mathop{\rm SU}(2)$ space. Now the relative choice of this projection
matrix and $Q$ in~(\ref{vpr-matrixQ}) matters. If they anticommute, the
choice that has been taken in~\cite{vpr-Bagger,vpr-Zucker}, then $g$ does
not change when one crosses the brane. If they commute, as
in~\cite{vpr-Pomarol,vpr-Polish}, then $g$ jumps over the brane. And the
latter is what we will take further.

After these general remarks, we come to \textbf{step~1}. We thus consider
the action of supergravity coupled to $n$ vector
multiplets~\cite{vpr-GST}. The fields are
\begin{equation}
    e_\mu ^a\,,\ \psi _\mu ^i\,,\ A_\mu ^I\,,\ \varphi ^{x}\,,\
  \lambda ^{ix}\,,
 \label{vpr-bulkfields}
\end{equation}
i.e.\ the graviton, gravitini, $n+1$ gauge fields ($I=0,1,\ldots ,n$),
including the graviphoton, $n$ scalars ($x=1,\ldots ,n$), and $n$ doublets
of spinors. The scalars describe a manifold structure that has been
called very special geometry~\cite{vpr-brokensi}. That geometry, and the
complete action, is determined by a symmetric tensor $C_{IJK}$. The
scalars are best described as living in an $n$-dimensional scalar manifold
embedded in an $(n+1)$-dimensional space. $h^I$ are the coordinates of
this larger space. The submanifold is defined by an embedding condition
such that the $h^I$ as functions of the independent coordinates $\varphi
^x$ should satisfy
\begin{equation}
  h^{I}(\varphi )h^{J}(\varphi)h^{K}(\varphi )   C_{IJK}=1\,.
 \label{vpr-embeddingC}
\end{equation}
The metric and all relevant quantities of this bulk theory is thus so far
only dependent on $C_{IJK}$.

Then we add the gauging of a $\mathop{\rm {}U}(1)$ group. That means that
we take a linear combination of the vectors as gauge field for this
$R$-symmetry. The linear combination is defined by real constants $V_I$:
\begin{equation}
  A_\mu ^{(R)}\equiv V_I A_\mu ^I\,.
 \label{vpr-AmuR}
\end{equation}
The action and the transformation laws are then modified by terms that
all depend on $\Blue{g Q^i{}_j}$.

In \textbf{step~2}, the coupling constant $g$ is replaced by a coupling
field $G(x)$. In the G\"{u}naydin--Sierra--Townsend (GST) action, the
coupling constant appears up to terms in $g^2$. We thus replace
\begin{equation}
 S_{GST}(\Blue{g})=S_0 +\Blue{g} S_1 + \Blue{g^2} S_2\ \Rightarrow\
S_{GST}(\Blue{G(x)})=S_0 +\Blue{G(x)} S_1 + \Blue{G(x)^2} S_2\,.
 \label{vpr-SGSTgG}
\end{equation}
Another term is added to the bulk action that forces $G(x)$ to be a
constant, using a Lagrange-multiplier 4-form $\Purple{A_{\mu \nu \rho
\sigma }}$:
\begin{eqnarray}
 S_{\rm bulk} &=&  S_{GST}(\Blue{G(x)})+ \int {\rm d}^5x\,e\,
 \frac{1}{4!} \varepsilon ^{\mu \nu
\rho \sigma\tau }  \Purple{A_{\mu \nu \rho \sigma }} \partial _\tau  \Blue{G(x)} \nonumber\\
 &=&  S_0 -\int {\rm d}^5x\, e\, V
- \int {\rm d}^5x\,e\, \Purple{\hat{F}(x)}  \Blue{G(x)} +\mbox{fermionic
terms.} \label{vpr-Sbulk}
\end{eqnarray}
In the second line, the terms have been reordered. The potential $V$
originates from $S_2$ in (\ref{vpr-SGSTgG}), and leads to the potential
\begin{equation}
  V=-6 \Blue{G^2} \left[W^2- \frac34\left( {\partial W\over \partial
  \varphi^x}\right)^2\right]\,,\qquad W  \equiv   \sqrt{\ft23}h^I V_I\,,
 \label{vpr-potBulk}
\end{equation}
where the linear combination $W$ appears, analogous to (\ref{vpr-AmuR}).
The third term in (\ref{vpr-Sbulk}) appears from integrating by part the
term with the Lagrange multiplier, leading to the flux
\begin{equation}
  \Purple{\hat{F}}\equiv \ft1{4!}e^{-1} \varepsilon ^{\mu \nu \rho \sigma \tau
}\partial _\mu \Purple{A_{\nu \rho \sigma \tau }}
+\mbox{covariantization.}
 \label{vpr-hatFlux}
\end{equation}
The covariantization terms come from $S_1$ in (\ref{vpr-SGSTgG}). This
method of describing a constant using a $(D-1)$-form is in fact an old
method that was already used in~\cite{vpr-Aurilia:1980xj}.

It is easy to understand how supersymmetry is preserved. Indeed, the GST
action is known to be invariant:
\begin{equation}
  \delta(\epsilon ) S_{GST}(\Blue{ g})=0\,.
 \label{vpr-GSTisinv}
\end{equation}
Therefore, the only non-invariance for $S_{GST}(\Blue{G(x)})$ appears, if
we define $\delta (\epsilon )\Blue{G}=0$, from the $x$-dependence of
$G(x)$. It is thus proportional to its spacetime derivative
\begin{equation}
\delta(\epsilon ) S_{GST}(\Blue{ G(x)})=B^\mu\, \partial _\mu
\Blue{G(x)}\,,
 \label{vpr-deltaSGSTG}
\end{equation}
where $B^\mu$ is some expression of the other fields and parameters, whose
exact form is not important for the argument here. One immediately sees
then that invariance of (\ref{vpr-Sbulk}) is obtained by defining the
transformation law of the 4-form as
\begin{equation}
  \delta(\epsilon ) \frac{1}{4!} \varepsilon ^{\mu \nu
\rho \sigma\tau } \Purple{A_{\mu \nu \rho \sigma }}=B^\tau = e\left[
-{\rm i} \ft32\overline{\psi }{}_\mu ^i\gamma ^{\mu \tau  }\epsilon
  ^j W-\overline{\psi }{}_\mu^i\gamma ^{\mu \tau \rho  }\epsilon
  ^j A^{(R)}_\rho +\ft32\overline{\lambda }{}^i_x W^{,x}
  \gamma ^\tau  \epsilon   ^j\right] Q_{ij}\,,
 \label{vpr-deltaA}
\end{equation}
where we gave also the explicit form for our case. However, it is clear
that the method is also valid in other theories.

\begin{figure}
\begin{center}
\leavevmode \epsfxsize=2.5in
 \epsfbox{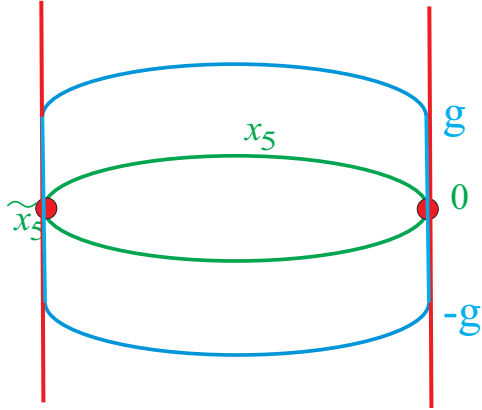}
\caption{\it The coupling constant $g$ jumps at $x_5=0$ and at $x_5=\tilde
x_5$. \label{vpr-fig:jumpg}}
\end{center}
\end{figure}

\textbf{Step~3} introduces the brane action, such that the total action is
\begin{equation}
   S_{\rm new} =  S_{\rm bulk}  + \Purple{S_{\rm brane}}\,.
 \label{vpr-Sbulk+brane}
\end{equation}
The brane action has the form
\begin{eqnarray}
   S_{\rm brane}&=&-2 \Blue{g }\int {\rm d}^5x\,\left( \ForestGreen{\delta (x^5)-\delta
(x^5-\tilde x^5 )}\right)
  \left( e_{(4)} 3 W +\ft1{4!} \varepsilon ^{{\underline{\mu}} {\underline{\nu}} {\underline{\rho}} {\underline{\sigma}} }\Purple{A_{{\underline{\mu}} {\underline{\nu}} {\underline{\rho}}
  {\underline{\sigma}} }}\right)\nonumber\\ &=& S_{brane,1} - S_{brane,2}\,.
 \label{vpr-Sbrane}
\end{eqnarray}
Underlined indices refer to the values in the brane directions:
$\underline{\mu }=0,1,2,3$. The action is presented as an integral over 5
dimensions, but the delta functions imply that it is a four-dimensional
action for each brane separately. The action of each brane consists of a
Dirac--Born--Infeld (DBI) term and a Wess--Zumino (WZ) term. However,
both parts depend only on the pullback of the bulk fields to the branes.
There are no fields living on the brane. The function $W$ appears in the
DBI term, and plays the role of the central charge of the brane. But most
importantly, the 4-form Lagrange multiplier appears in the WZ term, and
this thus modifies its field equation. The new field equation is
\begin{equation}
  \partial _5 \Blue{G(x^5)}= 2 \Blue{g}\left(\ForestGreen{ \delta (x^5)
-\delta (x^5-\tilde x^5)} \right)\,,
 \label{vpr-newFELm}
\end{equation}
and leads to the solution (taking into account the cyclicity condition)
\begin{equation}
  \Blue{G(x)}=\Blue{g}\,\ForestGreen{\varepsilon (x^5)}\,.
 \label{vpr-solGg}
\end{equation}
The function $\ForestGreen{\varepsilon (x^5)}$ jumps as well at $x^5=0$
as at $x^5=\tilde x^5$, see figure~\ref{vpr-fig:jumpg}.
It is clear from this picture that we need the second brane. Indeed, one
has to come back to the original value of $g$, in order that total
derivatives in $x^5$ do not contribute to the action. The flux, which is
determined by the field equation of $G(x)$, is
\begin{equation}
\Purple{\hat{F}} = 12\Blue{ G} \left[ W^2- \frac{3}{4}\left({\partial
W\over
\partial \varphi^x}\right)^2\right] +\mbox{fermionic terms.}
\end{equation}
The overall factor changes when crossing each brane due to
(\ref{vpr-solGg}). These jumps imply that the \emph{wall acts as a sink
for the fluxes}.

That supersymmetry is still preserved by the addition of the brane is
less obvious and is the non-trivial part of the construction. It turns
out that the supersymmetry is preserved thanks to the projections. One
finds (indices $m$ are tangent space indices in brane directions)
\begin{eqnarray}
  \delta S_{\rm brane}&=-3\Blue{g}\int d^5x\left(\ForestGreen{ \delta (x^5)
-\delta (x^5-\tilde x^5)} \right)e_{(4)} &
  \left[ W   \bar \epsilon ^i\gamma ^m e_m^{\underline{\mu}} \left(
  \psi _{{\underline{\mu}} i} - {\rm i} \gamma _5 \Blue{Q_{ij}}\psi _{\underline{\mu}}^j \right)+
  \right.\nonumber\\ && \left.\ \
  +\, W_{,x}
  \bar \epsilon  ^i\left( {\rm i}\lambda _i^x - \gamma _5 \Blue{Q_{ij}}
  \lambda ^{xj}
\right) \right].
\end{eqnarray}
The combinations of the gravitino and the gauginos that are in brackets
are the components that are odd under the $\mathbb{Z}_2$ projection, and
thus vanish on the brane. This leads to the invariance. Remark that in
each case one of the two terms comes from the DBI (mass) term and the
other from the WZ (charge) term. This therefore determines the relative
weight of the two terms, and is the mass $=$ charge relation, that says
that the brane is BPS. We thus see, indeed, that the brane action is
separately invariant. Note, that if we would not use (or eliminate) the
Lagrange multiplier, then this would relate bulk and brane, and only the
sum would be invariant.
\section{The background: BPS solutions}
We consider solutions with a warped metric, i.e.
\begin{equation}
  {\rm d} s^2= \ForestGreen{a^2(x^5)}\, {\rm d} x^{\underline{\mu}} {\rm d} x^{\underline{\nu}} \eta_{{\underline{\mu}}{\underline{\nu}}}
   + \ForestGreen{({\rm d} x^5)^2 }\,.
 \label{vpr-ds2warped}
\end{equation}
The energy density for solutions that depend only on $x^5$ is
\begin{eqnarray}
E(x^5)&=&-6a^2 a^{\prime 2} +\ft12 a^4(\varphi^{x\prime} )^2
+a^4V-\ft1{4!}\varepsilon ^{\mu \nu \rho \sigma 5 }\Purple{A_{\mu \nu
\rho \sigma }}  \Blue{G'}+ \nonumber\\ && +\, 2 \Blue{g} \left(
\ForestGreen{ \delta (x^5)-\delta (x^5-\tilde x^5 )}\right)
  \left( 3  a^4 W  + \ft{1}{4!}
  \varepsilon ^{{\underline{\mu}} {\underline{\nu}} {\underline{\rho}} {\underline{\sigma}} }\Purple{A_{{\underline{\mu}} {\underline{\nu}} {\underline{\rho}}
  {\underline{\sigma}} }}\right),
\end{eqnarray}
where the prime denotes a derivative w.r.t. $x^5$. The first three terms
come from the GST action, the last one on the first line from the term
that we added with the Lagrange multiplier. The second line comes from
the brane action. For this type of brane actions, one can rewrite it
using squares and total derivatives:
\begin{eqnarray}
E&=& \frac12  a^4 \left \{ \left[  \varphi^{x\prime} - 3 \Blue{G}
 W^{,x}  \right]^2 - 12 [  {a'\over a} + \Blue{G} W]^2\right \} + 3   [
a^4 \Blue{G} W]'+ \nonumber\\
&& +\, \left[ 2 \Blue{g} \left(  \ForestGreen{\delta (x^5)-\delta
(x^5-\tilde x^5 )}\right)-\Blue{G'}\right]
  \left( 3 a^4 W  + \ft{1}{4!}\varepsilon ^{{\underline{\mu}} {\underline{\nu}}
  {\underline{\rho}} {\underline{\sigma}} }\Purple{A_{{\underline{\mu}}
  {\underline{\nu}} {\underline{\rho}}
  {\underline{\sigma}} }}\right).
\end{eqnarray}
The expression in square brackets in the second line is the field
equation of the Lagrange multiplier, and this line can thus be omitted.
The last term of the first line is a total derivative in $x^5$ and thus
also does not contribute to the energy due to the continuity of the
fields. The vanishing of the squared terms gives thus the minimum of the
energy, and this minimum is even zero, as the zero energy of a closed
universe. The BPS conditions are thus
\begin{equation}
  \varphi^{x\prime} = 3 \Blue{G}\,  W^{,x}\,,\qquad {a'\over a} =- \Blue{G}\,
  W\,.
 \label{vpr-stabeqs}
\end{equation}
These equations are also called stabilization equations. These equations are
important to investigate the preserved supersymmetries. The
transformations of the fermions are
\begin{eqnarray}
 \delta (\epsilon )\lambda _i^x & = & -{\rm i}\ft12 \gamma _5\varphi
 ^{x\prime}\epsilon _i
 -\ft32 \Blue{GQ_{ij}} W^{,x} \epsilon ^j\, , \nonumber\\
 \delta (\epsilon )\psi _{{\underline{\mu}} i} & = &\partial _{\underline{\mu}} \epsilon _i
 + \ft12 \delta _{\underline{\mu}}^m \gamma
 _m\left( a'\gamma _5\epsilon _i+{\rm i} a \Blue{GQ_{ij}} W \epsilon ^j\right)\, ,
  \nonumber\\
 \delta (\epsilon )\psi _{5i} &=& \epsilon '_i+\ft12{\rm i} \Blue{GQ_{ij}} W\gamma _5\epsilon
 ^j\,. \label{vpr-transfoFermions}
\end{eqnarray}
To solve these, we split the supersymmetries in their even and odd parts:
\begin{equation}
 \epsilon _i  = \epsilon _i^+ + \epsilon _i^-\,, \qquad
 \epsilon _i^\pm  =  \ft12
 \left(\epsilon _i \pm {\rm i} \gamma _5\Blue{Q_{ij}}\epsilon ^j\right)=
 \pm {\rm i}\gamma _5\Blue{Q_{ij}}\epsilon ^{\pm j}\,.
\end{equation}
The vanishing of the last transformation of (\ref{vpr-transfoFermions})
determines the $x^5$ dependence of both parts. We have $ \epsilon _i^\pm
=a^{\pm 1/2}\epsilon _i^\pm (x^{\underline{\mu}})$. The transformations
of the other components of the gravitino then determines the dependence
on the other four spacetime variables. This gives the general solution,
\begin{equation}
\epsilon _i= a^{1/2}\epsilon _i^{+(0)}
  +a^{-1/2}\left(1-\frac{a'}{a}x^{\underline{\mu}} \gamma _{\underline{\mu}}\gamma _5\right)
  \epsilon_i^{-(0)}\,,
 \label{vpr-epsilongen}
\end{equation}
as function of $\epsilon_i^{\pm (0)}$, which are constant spinors with
each only 4 real components. There remains the transformations of the
gaugino, which lead to
\begin{equation}
  \varphi ^{x\prime}\epsilon_i^{-(0)}=0\,.
 \label{vpr-Twoposs}
\end{equation}
This leaves two possibilities. The first factor can be zero, which
implies that we have constant scalars. In that case 8 Killing spinors
survive. The other possibility allows non-constant scalars. Then the
second factor should be zero, and this thus eliminates 4 supersymmetries.
There remain 4 Killing spinors, $\epsilon_i^{+(0)}$, which are the 4 that
are non-vanishing also on the brane.

We consider both possibilities. First, let us look at the situation with
\emph{fixed scalars}. The BPS equations are then
\begin{equation}
  (\varphi^y)'  = 0 \,, \qquad
\left({\partial W \over \partial \varphi^x}\right)_{\rm crit} =0 \,,
 \qquad\frac{a'}{a} = -  g \varepsilon (x^5)  W\,.
 \label{vpr-BPSfixed}
\end{equation}
The constancy of $W$ is translated by formulae of very special geometry
in a `supersymmetric attractor equation'
\begin{equation}
   C_{IJK}\bar h^J \bar h^K  = q_I \,,\qquad \bar h^K \equiv  \sqrt{W_{crit}}
   h^K\,,\qquad  q_I\equiv \sqrt{\ft23} V_I\,.
 \label{vpr-Attr1}
\end{equation}
This equation is well-known from black-hole physics~\cite{vpr-FKS}. A
solution gives rise to a metric of the form
\begin{equation}
  {\rm d} s^2= e^{-2 \Blue{g} W_{\rm crit}|\ForestGreen{x^5}|}
   {\rm d} x^{\underline{\mu}} {\rm d} x^{\underline{\nu}} \eta_{{\underline{\mu}}{\underline{\nu}}} + \ForestGreen{({\rm d}
   x^5)^2}\,,
   \qquad \mbox{or}\qquad a=e^{-2 \Blue{g} W_{crit}|\ForestGreen{x^5}|}\,.
 \label{vpr-metricfixed}
\end{equation}
In this case, the negative-tension brane can be pushed to infinity.
Indeed, there is no obstruction as $a$ never vanishes.

To consider supersymmetric domain walls with \emph{non-constant scalars},
we use another coordinate, $y$, such that $\ForestGreen{{\partial\over
\partial x^5}= a^2 {\partial \over \partial y}}$. The metric is then
\begin{equation}
  {\rm d} s^2= \ForestGreen{a^2(y)} {\rm d} x^{\underline{\mu}} {\rm d} x^{\underline{\nu}} \eta_{{\underline{\mu}}{\underline{\nu}}}
   + \ForestGreen{a^{-4}(y){\rm d} y^2}\,.
 \label{vpr-metricncsc}
\end{equation}
The stabilization equations take the form
\begin{equation}
  a^2\ForestGreen{ \frac{{\rm d}}{{\rm d} y}}{ \varphi^x}  =   3  \Blue{G(y)}    W^{,x} \,, \qquad
a  \ForestGreen{\frac{{\rm d}}{{\rm d} y}}a = -  \Blue{G(y)}\, W\,.
 \label{vpr-stabncsc}
\end{equation}
These $n+1$ equations are combined, using relations of very special
geometry, to
\begin{equation}
   \ForestGreen{{{\rm d}\over {\rm d} y}} ( C_{IJK}\tilde h^J \tilde h^K)  = - 2 \Blue{G(y)}q_I
    \qquad \mbox{where}\qquad  \tilde h^I \equiv \ForestGreen{a(y)} h^I\,,
 \label{vpr-combstabncsc}
\end{equation}
whose solutions are given in terms of harmonic functions $H_I(y)$:
\begin{equation}
  C_{IJK}\tilde h^J \tilde h^K  = H_I(y)=c_I - 2\Blue{g} q_I
  \ForestGreen{|y|}\,,
 \label{vpr-attr2}
\end{equation}
where $c_I$ are integration constants, while $q_I$ are the constants that
were introduced in the gauging ($V_I$ up to a normalization). They are
harmonic in the sense that
\begin{equation}
  \ForestGreen{{{\rm d}\over {\rm d} y}  {{\rm d}\over {\rm d} y} }H_I=
 -4\Blue{g} q_I [\ForestGreen{\delta(y) - \delta(y-\tilde y)}]\,.
 \label{vpr-harmonicH}
\end{equation}
The warp factor is
\begin{equation}
  \ForestGreen{a^2(y)}=h^IH_I\,.
 \label{vpr-warpncsc}
\end{equation}
In this case the distance between the branes is restricted. There can be
two types of restrictions:
\begin{enumerate}
  \item There can be fundamental restrictions due to the origin of the
  functions $h^I$. E.g.\ these are in various applications related to
  integrals over Calabi--Yau cycles. Their vanishing can put a restriction
  on the distance.
  \item The vanishing of the harmonic functions also puts a restriction.
  Indeed, these harmonic functions enter in the warp factor, which should
  be non-vanishing.
\end{enumerate}
In each case this restricts the distance to be smaller than a critical
distance
\begin{equation}
  \ForestGreen{|\tilde y|}< \ForestGreen{|y|_{\rm sing}}\,.
 \label{vpr-restrictDist}
\end{equation}
\section{Summary and outlook}
The RS scenario in 5 dimensions can be made supersymmetric despite the
singularities of the space. The action and transformation laws can be
obtained using a 4-form, such that bulk and brane are separately
supersymmetric. Supersymmetric solutions exist with fixed scalars or 1/2
supersymmetry.

Half of the supersymmetries vanish on the branes. Also the translation
generator in the fifth direction vanishes on the brane. That is how the
algebra can be realized. These algebraic aspects could still be clarified
further. Also the extension to hypermultiplets deserves further study.
The same mechanism could be applied to study 8-branes in $D=10$ and other
similar situations. It is furthermore an intriguing question how
supersymmetric matter can live on the branes.

\medskip
\section*{Acknowledgments.}

\noindent  This work was supported by the European Commission RTN
programme HPRN-CT-2000-00131, in which E.B. is associated with Utrecht
University. The work of R.K.  was supported by NSF grant PHY-9870115.

\end{document}